\documentclass[10pt, conference]{IEEEtran}
\IEEEoverridecommandlockouts
\usepackage{amsmath,amssymb,amsfonts}
\usepackage{bbm}
\usepackage{algorithmic}
\usepackage{graphicx}
\usepackage{textcomp}
\usepackage{xcolor}
\usepackage{dsfont}
\usepackage{physics}
\usepackage{comment}
\usepackage{hyperref}
\usepackage[acronym]{glossaries}
\def\BibTeX{{\rm B\kern-.05em{\sc i\kern-.025em b}\kern-.08em
    T\kern-.1667em\lower.7ex\hbox{E}\kern-.125emX}}

\counterwithin{equation}{section}
\newcounter{thm}

\newtheorem{lemma}[thm]{Lemma}

\newcommand{\eins}{\mathbbm{1}}

\newacronym{povm}{POVM}{Positive Operator Valued Measurement}
\newacronym{csi}{CSI}{Channel State Information}

\begin{document}

\title{Infinite-fold Asymptotic Quantum Advantage in Classical Optical Correlation Sensing\\
\author{\IEEEauthorblockN{Janis N\"otzel, \emph{Member, IEEE}, Pere Munar-Vallespir}
\textit{Emmy-Noether Group Theoretical Quantum Systems Design},\\
\textit{Technical University of Munich, Munich, Germany},\\
\textit{\{janis.noetzel,pere.munar\}@tum.de}}
\thanks{This work was financed by the DFG via grant NO 1129/2-1 and by the Federal Ministry of Education and Research of Germany via grants 16KISQ093, 16KISQ039 and 16KISQ077. The generous support of the state of Bavaria via Munich Quantum Valley, the NeQuS and the 6GQT project is greatly appreciated. Finally, the authors acknowledge the financial support by the Federal Ministry of Education and Research of Germany in the program “Souverän. Digital. Vernetzt.”. Joint project 6G-life, project identification number: 16KISK002. }}

\maketitle
\begin{abstract}
    We study the hypothesis testing problem of distinguishing between correlated thermal noise and uncorrelated thermal noise of the same average energy on $K$ detectors in asymptotic asymmetric hypothesis testing. We compare the performance of heterodyne or homodyne detection with classical post-processing, the most general quantum strategy (involving any arbitrary measurement) and a simple strategy involving a photonic chip and On-Off detection. When the average received energy per detector goes to zero, the photonic chip strategy asymptotically achieves the optimal decrease in the error, while heterodyne/homodyne measurements do not. Thus, we show that linear optics and On-Off measurement are enough to achieve better detection than classical methods when detecting correlations in thermal optical signals. 
\end{abstract}

\section{Introduction}
    Given $K$ received signals, the ability to quantify their correlations is of importance in a variety of domains, ranging from signal detection and processing in communications \cite{tensorDecompositionSignalProcessing} over medical signal analysis in important techniques such as electroencephalography, magnetoencephalography, or prediction of post-operative complications \cite{tensorDecompositionEEG,tensorDecompositionMEG,tensorDecompositionPredictionOfComplication} over to financial data analysis \cite{tensorDecompositionFinancialMarket}. 
    We refer the reader to \cite{KoBa09} for a more general overview.
    
    Mathematically, the topic is related to tensor decompositions - in fact, all the above references utilize a tensor structure at the heart of the so-called CANDECOMP/PARAFAC decomposition, which can fit the received signals to a known structure. In the case at hand, this particular structure is given by a Kruskal tensor, which takes for $K=3$ forms such as $     \sum_r\lambda_r W^{(1)}_r\otimes W^{(2)}_r\otimes W^{(3)}_r$
    in signal processing \cite{tensorDecompositionSignalProcessing} and  $\sum_xp(x)w_1(x_1|x)w_2(x_2|x)w_3(x_3|x)$
    in information theory \cite{deducingTruth}, where $w_i(x_i|x)$ are conditional probabilities. Interestingly, such systems have not been studied from the perspective of quantum theory so far. The hidden assumption that a joint piece "$i$" of information (an arbitrary quantum state) is transmitted into various modes $W^{(1)},W^{(2)},W^{(3)}$ or $w_1,w_2,w_3$ is suspicious from the perspective of quantum theory due to the no-cloning theorem \cite{noCloning}. However, this is a reasonable assumption in some scenarios, such as classical light sources. 
    
    Such emission may take place in several ways, out of which we focus on the particular one where the received state $\rho_E^{(K)}$ ($K\cdot E$ denoting the total received energy) equals 
    \begin{align}
        \rho^{(K)}_E &= \tfrac{1}{\pi \cdot E}\int e^{-|\alpha|^2/E}|\alpha\rangle\langle\alpha|^{\otimes K}d\alpha
    \end{align}
    and describes the emission of \emph{a single thermal light source} towards the $K$ detectors (we use $|\alpha\rangle\langle\alpha|$ to denote coherent states \cite{serafini}). The other hypothesis is then be given by 
    \begin{align}
        \rho^{\otimes K}_E &= \bigotimes_{i=1}^K\tfrac{1}{\pi \cdot E}\int e^{-|\alpha|^2/E}|\alpha\rangle\langle\alpha|d\alpha
    \end{align}
    which describes a scenario where each detector receives the thermal emissions of a different source. Note that we use the Glauber-Sudarshan $P$-representation, as it will simplify calculations in the following, and that both of these states are gaussian. For an introduction on phase space representations of gaussian states, see \cite{Olivares_2012}.

    Our model describes the application of quantum methods to the sensing of specific properties of classical fields, and provides an alternative perspective to the known sensing mechanisms based on Greenberger–Horne–Zeilinger states or squeezed light \cite{ghzMetrology,serafini}. From the application perspective, our work may be of relevance to the detection of hidden emitters pointing to the presence of adversarial parties in the context of physical layer security \cite{wiretapCaiWinterYeung}, where empirical validation of assumptions on channel models is critical. Some communication scenarios, such as satellite communication \cite{losses-sat-comms} or under-water communication \cite{under-water,amiri}, do as well motivate a low photon regime due to high loss and the absence of amplifiers. In such scenarios, the hypothesis testing task can be seen as an attempt by the receiver to identify if an incoming signal qualifies as background noise or has instead a specific pattern. 
    
    Leaving aside the motivating potential applications, we formalize the task to be carried out as one of asymmetric hypothesis testing to properly quantify performance for each strategy. In particular, we ask what the error exponent is in asymptotic asymmetric hypothesis testing when we desire to correctly detect the correlation of the signals with a fixed probability $1-\epsilon$ while minimizing the probability of falsely claiming the signals are not correlated (this probability is usually called the type two error). This problem is loosely related to the study of data transmission over quantum systems in the sense that, for $K=2$, the error exponent of the type two error equals the quantum mutual information of $\rho_{E}^{(2)}$. The exponents are given by quantum relative entropies, which are related to performance in other tasks such as quantum quickest change-point detection (QUSUM) \cite{john2025fundamentallimitsquickestchangepoint}\cite{Fanizza_2023} and sequential hypothesis testing \cite{Li_2022}\cite{seq_hyp_test}. So the quantum advantage we describe is not limited to hypothesis testing, even if we use this task as initial motivation. Our task is, in a sense, similar to quantum illumination \cite{Wilde_2017}, but we do not require entanglement. 
    
    For the particular scenario we studied, we prove the asymptotic superiority of the quantum detection method over its classical counterpart, which can be quantified by observing that the ratio of the respective error exponents scales as $1/E$ for $E\to0$. We focus on the low photon number, where we observe that the strategy that achieves the optimal performance is implementable with current technology. In particular, optimal chips of up to 20 modes have already been demonstrated  \cite{Taballione_2023}, and there is ongoing research on them for applications in AI \cite{zhu2025versatilesiliconintegratedphotonic}.
    
\section{Problem Statement}
The mathematical problem statement is taken from (quantum) hypothesis testing theory. For the two quantum states $\rho$ and $\sigma$, the task is to determine for $\epsilon>0$ the quantity
\begin{align}
	D_Q(\rho,\sigma):=\lim_{n\to\infty}-\frac{1}{n}\log\beta_{n,\epsilon}(\rho,\sigma), 
\end{align}
where $\beta_{n,\epsilon}(\rho,\sigma):=\min_{0\leq P\leq\mathbbm1^{\otimes n}}\{\tr P\sigma^{\otimes n}:\tr \left((\mathbbm{1}-P)\rho^{\otimes n}\right)\geq1-\epsilon\}$ is called the ``type two error''. Physically, we assume we can apply any measurement on $n$ copies of the state (either $\rho^{\otimes n}$ or $\sigma^{\otimes n}$), and then we have to decide which state we were given copies of. The exponent quantifies how fast the type-II error decreases while keeping the type-I error constant. This decay is exponential in the number of samples, and we are interested in the largest exponent that can be achieved. 

For finite dimensional systems, the error exponent $D_Q(\rho,\sigma)$ has been shown to equal the quantum relative entropy $D(\rho\|\sigma):=\tr(\rho(\log\rho-\log\sigma))$ in the seminal works \cite{hiaiPetz,ogawaNagaoka}. The more general case of $W^*$-algebras has also been studied in \cite{quantum_hyp_von_Neumann}, and for the particular case of Gaussian states, which includes our particular case, it was shown to also equal the quantum relative entropy in \cite{Spedalieri_2014}. We include a derivation of the result using finite-dimensional approximations for completeness. Similar problems have been studied in the classical setting. If instead the two hypotheses are given by classical multivariate normal distributions on $\mathbb R^n$, we may formulate the problem statement along the lines of the recent work \cite{lungu2024finitesampleexpansionsoptimalerror} as the calculation of 
\begin{align}
	D_C(\mathcal N_1,\mathcal N_2):=\lim_{n\to\infty}-\frac{1}{n}\log\alpha_{n,\epsilon}(\mathcal N_1,\mathcal N_2), 
\end{align}
where $\alpha_{n,\epsilon}(\mathcal N_1,\mathcal N_2):=\inf_{p_{Z|X}}\{e_1(p_{Z|X}):e_2(p_{Z|X})\leq\epsilon\}$, where $e_i(p_{Z|X})=\int p_{Z|X}(i|x)dp_i(x)$ and $p_i$ are the probability density functions of the multivariate normal distributions $\mathcal N_i$ . For a large class of distributions, \cite{lungu2024finitesampleexpansionsoptimalerror} proved a formula for $D_C$, which we use for $n$-variate normal distributions in the form $D_C(\mathcal N_1,\mathcal N_2)=D(\mathcal N_1\|\mathcal N_2)$, where $D(\mathcal N_1\|\mathcal N_2)=\int p_1(x^n)\log(p_1(x^n)/p_2(x^n))dx^n$ is the usual Kullback-Leibler divergence. Here and in the following  the logarithm $\log$ is taken with base $2$. 

\section{Results}

Our method of analysis is to compare the state-of-the-art technology with the hypothetical future quantum systems when the goal is to distinguish between the hypothesis $\rho=\rho_E^{(K)}$ and $\sigma=\rho_E^{\otimes K}$. Since our systems are optical states, they are defined on the infinite-dimensional Fock space $\mathcal F^{\otimes K}$, where $\mathcal F:=\mathrm{span}\{|n\rangle\}_{n\in\mathbb N}$, where $|n\rangle$ are the so-called \emph{number states} \cite{serafini}. To show the desired quantum advantage, we give an expression for $D_Q(\rho_E^{(K)},\rho_E^{\otimes K})$. This is then compared to an approach that qualifies as ``state of the art''. For this state of the art approach, we let the $K$ detectors perform either heterodyne or homodyne detection on their received signal before sending it forward to the processing center, which will perform classical post-processing. Both detection methods are well-established parts of today's communication infrastructure. In both cases, we then obtain multi-variate normal distributed random variables, which we write as $p_E^{(K)}$ when the correlated state $\rho_E^{(K)}$ is given and $p_E^{\otimes k}$ if the uncorrelated quantum state $\rho_E^{\otimes K}$ is given. We will later see that the probability density function $p_E^{(K)}$ is, for heterodyne detection, given by 
\begin{align}
	p_E^{(K)}(x^k&)= \tfrac{1}{\pi^{k+1} E}\int e^{-\tfrac{|\alpha|^2}{E}}e^{-\sum_{i=1}^K|\alpha-x_i|^2}d\alpha
\end{align}
and for homodyne detection by 
\begin{align}
	\tilde p_E^{(K)}(x^k&)= \tfrac{1}{\pi^{k+1} E}\int e^{-\tfrac{|\alpha|^2}{E}}e^{-\sum_i|\sqrt{2}\real(\alpha)-x_i|^2}d\alpha.
\end{align}
From classical results, we note that $D_{Het}(\rho_E^{(K)},\rho_E^{\otimes K}) = D(p_{E}^{(K)}\|p_E^{\otimes K})$ and $D_{Hom}(\rho_E^{(K)},\rho_E^{\otimes K}) = D(\tilde p_E^{(K)}\|\tilde p_E^{\otimes K})$. 
The corresponding marginal distributions $p_E,\tilde p_E$ and their $K$-fold products $p_E^{\otimes K},\tilde p_E^{\otimes K}$ are then derived in the usual way and allow us to state our main results as follows.
The error exponents are given by:
\begin{align}
\label{thm:error-exponents}
    D_Q(\rho_E^{(K)},\rho_E^{\otimes K}) &= K\cdot g(E) - f(1+2\cdot K\cdot E)\\
    D_{\scriptscriptstyle{\mathrm{Ht}}}(\rho_E^{(K)},\rho_E^{\otimes K}) &= \log\frac{(1+E)^K}{(1+K\cdot E)}\\
    D_{\scriptscriptstyle{\mathrm{Hm}}}(\rho_E^{(K)},\rho_E^{\otimes K})&=\frac{1}{2} \log\frac{(1+2E)^K}{(1+2K\cdot E)}\\
    D_{\scriptscriptstyle{\mathrm{PH}}}(\rho_E^{(K)},\rho_E^{\otimes K}) &= d(E,K) + (K-1) \log(1+E)
\end{align}
where $g$ is the Gordon function defined via $g(x):= (x+1)\log(x+1)-x\log x$ for $x>0$ and $g(0)=0$, $f(x):=\tfrac{x+1}{2}\log\tfrac{x+1}{2}-\tfrac{x-1}{2}\log\tfrac{x-1}{2}$ for $x>1$, $f(1)=0$ and $d(E,K)$ is the relative entropy $D(\tfrac{1}{1+KE}|| \tfrac{1}{1+E})$ between two Bernoulli random variables with parameters $\tfrac{1}{1+KE}$ and $\tfrac{1}{1+E}$. The exponents are, in order, the optimal quantum limit, achievable with heterodyne measurement, achievable with homodyne measurement and achievable with linear optics and On-Off measurement. 

From \cite{Banaszek_2020}, we are inclined to believe the interesting system behavior is in the regime $E\approx0$, which turns out to be true. In particular, by expanding for small energies we find that
\begin{align}
\label{thm:quantum-advantage}
    D(\rho_E^{(K)},\rho_E^{\otimes K})& =  K\cdot \ln(K)\cdot E + \mathcal O(E^2)\\
    D(p_{E}^{(K)}\|p_E^{\otimes K}) &= \tfrac{(K-1)\cdot K}{\ln(4)}\cdot E^2 + \mathcal O(E^3)\\
    D(\tilde p_E^{(K)}\|\tilde p_E^{\otimes K}) &=\tfrac{2(K-1)\cdot K}{\ln(2)}\cdot  E^2+\mathcal O(E^3)\\
    D_{\textit{PH}}(\rho_E^{(K)},\rho_E^{\otimes K}) &= K\cdot \ln(K)\cdot E + \mathcal O(E^2).
\end{align}
Thus we have $\lim_{E\to0}D(\rho_E^{(K)},\rho_E^{\otimes K})/D(p_{E}^{(K)}\|p_E^{\otimes K})=\infty$ and $\lim_{E\to0}D(\rho_E^{(K)},\rho_E^{\otimes K})/D(\tilde p_{E}^{(K)}\|\tilde p_E^{\otimes K})=\infty$. Finally, we note that the optimal scaling for low energy can be obtained using passive optics and On-Off detection employing a Hadamard interferometer \cite{Guha_2011}.

\begin{figure}
	\centering
	\includegraphics[width=.45\textwidth]{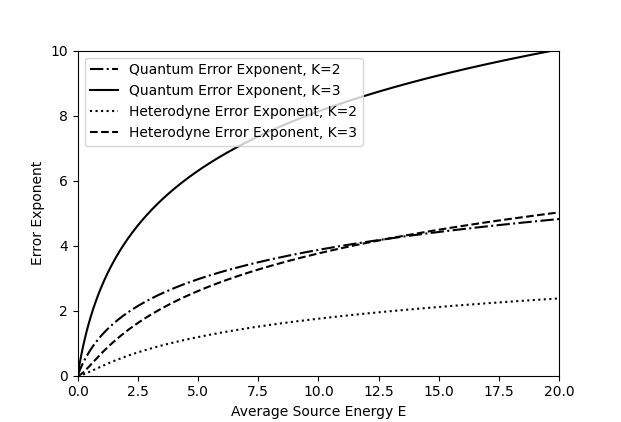}
	\caption{Depicted is the (unbounded) quantum advantage in the error exponent for low energy and the saturation for large energy values. As can be seen, access to a large enough number of observations at a high enough energy can allow the classical system to outperform a corresponding quantum system with access to fewer copies.  \label{fig:quantum-advantage}}
\end{figure}

Results are illustrated in Figure \ref{fig:quantum-advantage}. The general notation for quantum systems is borrowed from \cite{serafini}. Note that, while all exponents go to 0 as energy is reduced, the optimal quantum exponent and the one using a Hadamard interferometer go to 0 linearly in $E$ while the heterodyne and homodyne do so quadratically. So for the same number of detectors, we get a $\sim 1/E$ advantage in the exponent.

\section{Proofs and calculations}
	We start by proving the formula for the relative entropy $D(p_E^{(K)}\|p_E^{\otimes K})$ in Subsection \ref{subsec:heterodyne}, from which the one for $D(\tilde p_E^{(K)}\|\tilde p_E^{\otimes K})$ follows with minor modifications in Subsection \ref{subsec:homodyne}, and finally show that $D(\rho_E^{(K)}\|\rho_E^{\otimes K})$ is given by the relative entropy in Subsection \ref{subsec:quantum}.
	\subsection{Analysis State of the Art: Heterodyne Detection\label{subsec:heterodyne}}
	For the state-of-the-art version of the hypothesis testing problem, we assume the $K$ detectors perform heterodyne or homodyne detection on their received signal before sending it forward to the processing center. In both cases, they will obtain multi-variate normal distributed random variables $p_E^{(K)}$ in case the correlated state $\rho_E^{(K)}$ is given and $p_E^{\otimes k}$ in case the uncorrelated quantum state $\rho_E^{\otimes K}$ is given. In such cases, the relative entropy $D(p_E^{(K)}||p_E^{\otimes K})$ can be calculated based on a well-known identity. The probability density of a multi-variate normal distribution $\mathcal N(\mu,\Sigma)$ on $n$ modes can be written as 
	\begin{align}
		\mathbb R^n\ni t \to (2\pi)^{-n/2}\det(\Sigma)^{-1/2}e^{-(t-\mu)^T\Sigma^{-1}(t-\mu)}
	\end{align}
	where $\mu\in\mathbb R^n$ is the mean and the $n\times n$ matrix $\Sigma$ the variance of the distribution. For two such distributions $\mathcal N_1,\mathcal N_2$ with means $\mu_1,\mu_2$ and variances $\Sigma_1,\Sigma_2$ we have \cite[Chapter 1]{statisticalInferenceBook}
	\begin{align}\label{eqn:relative-entropy-for-multivariate-normal}
		  D(\mathcal N_1\|\mathcal N_2) = \tfrac{1}{2\ln(2)}(D+M+T),
	\end{align} 
        where $D:=\ln\det(\Sigma_2)-\ln\det(\Sigma_1)$, $M :=(\mu_2-\mu_1)^T\Sigma_2^{-1}(\mu_2-\mu_1)$ and $T:=\Tr{\Sigma_2^{-1}\Sigma_1-\eins_n}$. Normal distributions enter our analysis via heterodyne detection applied to a coherent state $|\alpha\rangle\langle\alpha|$, leading \cite{weedbrook} to a distribution of measurement results $x\in\mathbb C$ as 
	\begin{align}
		  p(x|\alpha)=H_{x}(|\alpha\rangle\langle\alpha|)=\tfrac{1}{\pi}e^{-|\alpha-x|^2}.
	\end{align}
	If the source is measured using $K$ heterodyne detectors, the probability density for receiving result $x^k=(x_1,\ldots,x_k)$ is
	\begin{align}
		  p^{(K)}(x^k&) 
		= \tfrac{1}{\pi E}\int e^{-\tfrac{|\alpha|^2}{E}}\prod_{i=1}^K H_{x_i}(|\alpha\rangle\langle\alpha|)d\alpha\\
		&= \tfrac{1}{\pi^{k+1} E}\int e^{-\tfrac{|\alpha|^2}{E}}e^{-\sum_{i=1}^K|\alpha-x_i|^2}d\alpha.\label{eq:joint-distribution}
	\end{align}
	Writing $x_i=x_{i1}+\mathbbm{i}x_{i2}$, the integration over the complex variable $\alpha=a+\mathbbm{i}b$ can be carried out based on the formula
	\begin{align}
		\int e^{-\tfrac{a^2}{E}-\sum_i(a-t_i)^2}da =C_{K,E}e^{-\sum_it_i^2+\tfrac{E(\sum_it_i)^2}{1+ K\cdot E}}\label{eq:integral-identity}
	\end{align}
	where summations run from $1$ to $K$ and $C_{K,E}:=\sqrt{\tfrac{\pi E}{1+K\cdot E}} $.
	By using the identity $|\alpha|^2=\real(\alpha)^2+\imaginary(\alpha)^2$ we conclude from \eqref{eq:joint-distribution} that the distribution of the complex measurement results $x^k$ is actually independent between their real and their imaginary parts: Defining
	\begin{align}
		q^{(K)}(t^k) = \sqrt{\tfrac{1}{\pi^{k+1} E}}\int e^{-\tfrac{a^2}{E}}e^{-\sum_{i=1}^K(a-t_i)^2}da\label{eq:real-distribution}
	\end{align}
	and using our convention $x_i=x_{i1}+\mathbbm{i}x_{i2}$ again we get
	\begin{align}
		q^{(K)}(x^k&)= q^{(K)}(\real{x^k})\cdot q^{(K)}(\imaginary{x^k})
	\end{align}
	where real- and imaginary parts of a vector are taken per component. Since $D(r\otimes s\|u\otimes v)=D(r\|u) + D(s\|v)$ we know that we can proceed to evaluate only the real part without loss of generality. We thus set $t^k:=\real{x^k}$ and proceed by calculating the matrix $\Sigma_{q^{(K)}}^{-1}$ corresponding to $q$, whose entries we can based on \eqref{eq:integral-identity} identify as:
	\begin{align}
		(\Sigma_{q^{(K)}}^{-1})_{ij} := \frac{(1+E(K-1))\delta_{i,j}-(1-\delta_{i,j})E}{1+K\cdot E}.
	\end{align}
	where and correspondingly according to Lemma \ref{lem:matrix-inverse} the inverse $\Sigma_{q^{(K)}}$ can be calculated to have a similar structure:
	\begin{align}
		(\Sigma_{q^{(K)}})_{i,j} = (1+E)\delta_{i,j}+E(1-\delta_{i,j}).
	\end{align}
	The marginals of $q^{(K)}$ all have identical distributions, and each of them is calculated to equal
	\begin{align}
		q(t) 
    		&= \sqrt{\tfrac{1}{2\pi(1+E)}}e^{-t^2/(1+E)}.
	\end{align}
	Since all correlations vanish, we have
	\begin{align}
		\Sigma^{-1}_{q^{\otimes K}} = \frac{1}{1+E}\eins.
	\end{align}
	It then holds
	\begin{align}
		\det(\Sigma_{q^{\otimes K}}) &= (1+E)^{K},\\
		\det(\Sigma_{q^{(K)}}) &= 1+K\cdot E\\
		\tr(\Sigma_{q^{\otimes K}}^{-1}\Sigma_{q^{(K)}})&= (1+E)^{-1}\tr(\Sigma_{q^{(K)}}) =K.
	\end{align}
	Using  \eqref{eqn:relative-entropy-for-multivariate-normal} and noting $M=0$ in our case, we get the result
	\begin{align}
		D(q_E^{(K)}\|q_E^{\otimes K}) &=2\cdot D(q^{(K)}\|q^{\otimes K})\\
		&= \log\det(\Sigma_{q^{\otimes K}})-\log\det(\Sigma_{q^{(K)}})\\
		&=\log(\tfrac{(1+E)^{K}}{1+KE}).
	\end{align}
	\subsection{Analysis State of the Art: Homodyne Detection\label{subsec:homodyne}}
	When homodyne detection is used, the conditional probability of getting the \emph{real-valued} measurement outcome $x$ given the state was $|\alpha\rangle\langle\alpha|$ is given by $p(x|\alpha)=\tfrac{1}{\sqrt{2\pi}}e^{-(x-\sqrt{2}\real(\alpha))^2/2}$ \cite[Eq. (11)]{Kawakubo_2010}. This is equivalent to studying the previous problem of heterodyne detection, but with only one quadrature and instead an energy which is increased by a factor of $2$. The relative entropy therefore, changes to 
	\begin{align}
		D(\tilde p_E^{(K)}\|\tilde p_E^{\otimes K}) 
		&= \tfrac{1}{2}\log(\tfrac{(1+2E)^{K}}{1+2KE}).
	\end{align}
    \subsection{Quantum Approach\label{subsec:quantum}}
        To prove the desired lower bound, we first have to deal with the problem that the system of interest is infinite-dimensional, and show that for the specific case treated here, we have $D_Q(\rho_E^{(K)},\rho_E^{\otimes K})\geq D(\rho_E^{(K)}\|\rho_E^{\otimes K})$ (hypothesis testing bound). Afterwards, we proceed by noting that $D(\rho_E^{(K)}\|\rho_E^{\otimes K}) = K\cdot S(\rho_E)-S(\rho_E^{(K)})$ and then calculating the two respective entropies. 
        \paragraph{Functional dependence of the exponent on the states}
            For convenience, we base our work on \cite{bjelakovic2012quantumsteinslemmarevisited}, which provides near-optimal tests even when the Hilbert space size grows logarithmically with the block length. We benefit from the fact that all involved systems have bounded energy, which allows us to proceed as follows: We truncate each involved system at finite energy, and use the optimal projection for the resulting finite-dimensional systems. By controlling the approximation error, we can utilize the continuity of entropy and the specific structure of our hypothesis testing problem to arrive at the desired result. Using Lemma \ref{lem:effective-dimension-for-phase-randomized-coherent-state}, we have with $T_N:=\sum_{n=0}^N|n\rangle\langle n|$
	\begin{align}
		\tr(T_N\rho_E)\geq1-2^{-N}\label{eqn:trace-bound}
	\end{align}
	for large enough $N$. We conclude that for a suitable $\kappa_1>0$:  
	\begin{align}
		\min\{\tr(T_N^{\otimes K}\rho_E^{\otimes K}),\tr(T_N^{\otimes K}\rho_E^{(K)})\}\geq1-2^{-\kappa_1 N}.
	\end{align}
	Define $\mathcal N(\rho):=T_N\rho T_N + \tr((\eins-T_N)\rho)|0\rangle\langle0|$ and
	\begin{align}
		\tilde\rho_E^{(K)}:=\mathcal N^{\otimes K}(\rho_E^{(K)}),\qquad \tilde\rho_E^{\otimes K}:=\mathcal N(\rho_E)^{\otimes K}.
	\end{align}
	Then from \eqref{eqn:trace-bound} we know that for some $\kappa_2>0$ 
	\begin{align}\label{eqn:norm-bound}
		\max\{\|\tilde\rho_E^{(K)}-\rho_E^{(K)}\|_1, \|\tilde\rho_E^{\otimes K}-\rho_E^{\otimes K}\|_1\}\leq2^{-\kappa_2\cdot N}.
	\end{align}	
	According to \cite[Eqn. (17)]{bjelakovic2012quantumsteinslemmarevisited} there is for every $\delta>0$ a sequence $(P_n)_{n\in\mathbb N}$ of non-negative operators so that for every $\epsilon>0$ there exists an $N_0\in\mathbb N$ such that for all $n\geq N_0$
	\begin{align}
		\tr(P_n\tilde\rho_E^{(K)})&\geq 1-\epsilon\\
		\tr(P_n\tilde\rho_E^{\otimes K})&\leq 2^{-n\cdot(D(\tilde\rho_E^{(K)}\|\tilde\rho_E^{\otimes K})-\delta)}.
	\end{align}
        Let $\{K_i\}_i$ be the Kraus representation of $\mathcal N$. Then we have, for every quantum state $\sigma$ and non-negative matrix $X$,
        \begin{align}
            \tr(X\mathcal N(\sigma))= \textstyle\sum_i\tr(XK_i\sigma K_i^\dagger)= \tr(\textstyle\sum_iK_i^\dagger XK_i\sigma). \nonumber
        \end{align}
        In our particular situation, $P_n':=\sum_iK_i^\dagger P_nK_i$ thus satisfies
        \begin{align}
		\tr(P'_n\rho_E^{(K)})&\geq 1-\epsilon\label{eqn:detection-performance-of-P'_n}\\
		\tr(P'_n\rho_E^{\otimes K})&\leq 2^{-n\cdot(D(\tilde\rho_E^{(K)}\|\tilde\rho_E^{\otimes K})-\delta)}.
	\end{align}
        Since $D(\tilde\rho_E^{(K)}\|\tilde\rho_E^{\otimes K})=K\cdot S(\tilde \rho_E)-S(\tilde \rho_E^{(K)})$ and since       $\tr(\hat n\tilde\rho_E)\leq\tr(\hat n\rho_E)$ and $\tr(\hat n_K\tilde\rho_E^{(K)})\leq\tr(\hat n_K\rho_E^{(K)})$ where $\hat n$ and $\hat n_K$ are the photon number operators for $1$ and $K$ modes, respectively, and the original ensembles have finite energy, we may utilize the fact that for such states the entropy is continuous in $\|\cdot\|_1$ (see \cite{Winter2015} and \cite{Becker2021}), so that \eqref{eqn:norm-bound} and the choice $N=\log(n)/\kappa_3$ for a suitably chosen $\kappa_3>0$ implies
	\begin{align}
		  D(\tilde\rho_E^{(K)}\|\tilde\rho_E^{\otimes K})\to D(\rho_E^{(K)}\|\rho_E^{\otimes K}),
	\end{align} 
        and at the same time $\tr(P_n\tilde\rho_E^{(K)})\geq 1-\epsilon-\tfrac{1}{n}$. Note that the speed of convergence in \cite{bjelakovic2012quantumsteinslemmarevisited} is tied to the variance $V_N$ of the function $-\log\mu_K$ defined via $\mu_K(n^k):=\langle n^k,\tilde \rho_E^{\otimes K}n^K\rangle$ with respect to the discrete random variable $N^K$ defined via $\tilde s_N(n^K):=\mathbb P(N^K = n^k)=\langle n^k,\tilde \rho_E^{(K)}n^k\rangle$. For $N=\infty$ it holds $V_\infty=E(1+E)\log^2((1+E)/E)$. Due to the structure of $s_N$ one can show that $V_N\leq V(1+2^{-\kappa_4\cdot N})$. Thus the central limit theorem for discrete random variables guarantees \eqref{eqn:detection-performance-of-P'_n}, as in \cite{bjelakovic2012quantumsteinslemmarevisited}.
        By construction of $\tilde\rho_E^{(K)}$ and $\tilde\rho_E^{\otimes K}$, this implies the inequality
	\begin{align}
		  D_Q(\rho_E^{(K)},\rho_E^{\otimes K})\geq D(\rho_E^{(K)}\|\rho_E^{\otimes K}).
	\end{align} 
Now we show that this is a lower and an upper bound, so it is the optimal value. We show this by using the classical version of Stein's lemma, and we see that the measured relative entropy is upper bounded by the quantum relative entropy. We denote the measured relative entropy as $D_{\mathcal{M}}(\rho,\sigma) = \sup_{M} D(P_{M, \rho}|| P_{M, \sigma})$ where $P_{M, \rho}$ is the classical probability distribution of outcomes of $\rho$ under the POVM $M$. The optimization is over all possible POVMs. From the classical result, we know that after applying the measurement, we find 
\begin{align}
    D_{Q}(\rho || \sigma) \leq \sup_{n}\frac{1}{n} D_{\mathcal{M}}(\rho^{\otimes n} || \sigma^{\otimes n})
\end{align}
and we are taking a supremum over $n$ as we allow for collective measurements over $n$ copies. Finally, we note that for all states (with finite energy)  $\tfrac{1}{n}D_{\mathcal{M}}(\rho^{\otimes n} || \sigma^{\otimes n}) \leq \tfrac{1}{n}D(\rho^{\otimes n}||\sigma^{\otimes n}) = D(\rho||\sigma)$, where the inequality comes from the data processing inequality \cite{Lindblad1975} and the equality comes from the additivity of the relative entropy for product states. These are fundamental and well-known properties of the relative entropy that also hold in infinite-dimensional systems.  
	\paragraph{Computation of Entropy}
	To perform calculations, we use the conventions as described in \cite{weedbrook} with the small add-on that we use subscripts like $\langle\hat A\rangle_\rho$ to indicate that the expectation of operator $\hat A$ is meant to be taken with respect to the state $\rho$. Using the convention of physical constants as in \cite{weedbrook,serafini}, we have canonical position- and momentum operators equal to $\hat q_k=\hat a_k+\hat a_k^\dagger$ and $\hat p_k=\mathbbm{i}(\hat a_k^\dagger-\hat a_k)$. The following expectation values are needed for the computation:
	\begin{align}
		\langle\hat p\rangle_\alpha  &=\tr(\hat p|\alpha\rangle\langle\alpha|) = 2\imaginary(\alpha)\label{eqn:alpha-expectation-of-p}\\
		\langle \hat q\rangle_\alpha &= \tr(\hat q|\alpha\rangle\langle\alpha|) = 2\real(\alpha)\label{eqn:alpha-expectation-of-q}\\
		\langle \hat p^2\rangle_\alpha &= \tr(\hat p^2|\alpha\rangle\langle\alpha|) = 1+4\real(\alpha)^2\label{eqn:alpha-expectation-of-p^2}\\
		\langle \hat q^2\rangle_\alpha &= \tr(\hat q^2|\alpha\rangle\langle\alpha|) = 1+4\imaginary(\alpha)^2\label{eqn:alpha-expectation-of-q^2}
	\end{align}
	Observations \eqref{eqn:alpha-expectation-of-p} and \eqref{eqn:alpha-expectation-of-q} imply
	\begin{align}
		\langle \hat p_k\rangle_{\rho_E^{(K)}}=\langle \hat q_k\rangle_{\rho_E^{(K)}} = 0\qquad \forall k\in \{1,\ldots,K\}.\label{eqn:mean-equals-zero}
	\end{align}
	To compute the covariance matrix of $\rho_E^{(K)}$ we use the integrals
	\begin{align}
		\tfrac{1}{\pi E}\int e^{-|\alpha|^2/E}\langle \hat p\rangle_\alpha\langle \hat p\rangle_\alpha d\alpha= 2E\label{eqn:expectation-of-p^2}\\
		\tfrac{1}{\pi E}\int e^{-|\alpha|^2/E}\langle \hat q\rangle_\alpha\langle \hat q\rangle_\alpha d\alpha= 2E\label{eqn:expectation-of-q^2}\\
		\tfrac{1}{\pi E}\int e^{-|\alpha|^2/E}\langle \hat p\rangle_\alpha\langle \hat q\rangle_\alpha d\alpha= 0\label{eqn:expectation-of-pq}.
	\end{align}
	Based on \eqref{eqn:mean-equals-zero} we can draw first conclusions on the structure of the covariance matrix $V^{(K)}$ of the state 
	$\rho_E^{(K)}$:
	\begin{align}
		(V^{(K)})_{ij}:=\tfrac{1}{2}\langle\hat x_i\hat x_j + \hat x_j\hat x_i \rangle_{\rho_E^{(K)}}.
	\end{align}
	If $i=j$ this implies $(V^{(K)})_{ii}=\langle\hat x_i\hat x_i \rangle_{\rho_E^{(K)}}$, and equations \eqref{eqn:alpha-expectation-of-p^2} and \eqref{eqn:alpha-expectation-of-q^2} tell us that $(V^{(K)})_{ii}=1+2E$ holds.
	
	If $i\neq j$ but both $i$ and $j$ and either $i$ is even and $j$ is odd or the other way around, we are looking at correlations between momentum and position of different subsystems, and equation \eqref{eqn:expectation-of-pq} implies that $(V^{(K)})_{ij}=0$ must hold. If instead $i$ and $j$ are both even or both odd, then equations \eqref{eqn:expectation-of-p^2} and \eqref{eqn:expectation-of-q^2} imply that $(V^{(K)})_{ij}=E$. Thus $V^{(K)}$ is a Toeplitz matrix defined by the vector $v^{2K}$ with entries $v_1=2E+1$ and $v_{2i}=0$ for $i=1,\ldots,K$ and $v_{2i+1}=2E$ for $i=1,\ldots,K-1$. The entropy of $\rho^{(K)}$ is calculated based on its symplectic eigenvalues $\nu_1,\ldots,\nu_K$, which equal the non-negative eigenvalues of the matrix $\mathbbm{i}\Omega V^{(K)}$ where 
	\begin{align}
		\Omega=\bigoplus_{k=1}^K\omega,\qquad\omega=\left(\begin{array}{ll}0&1\\-1&0\end{array}\right). 
	\end{align}
	Using the structure of our states and the fact that the thermal state $\rho_E$ has entropy $g(E)$ \cite{holevoBOOK}, we see that
	\begin{align}
		D(\rho_E^{(K)}\|\rho_E^{\otimes K}) 
		&= S(\rho_E^{\otimes K}) -S(\rho_E^{(K)}) \\
		&= K\cdot g(E) -\textstyle\sum_{i=1}^Kf(\nu_i).
	\end{align}        
	It remains to calculate the symplectic eigenvalues $\nu_1,\ldots,\nu_K$ to arrive at the desired result. For $K=2$, $V^{(2)}$ is given by
	\begin{align}
		V^{(2)} = 2\left(\begin{array}{llll}
			E+\tfrac{1}{2} & 0 & E & 0\\
			0 & E+\tfrac{1}{2} & 0 & E\\
			E & 0 & E+\tfrac{1}{2} & 0\\
			0 & E & 0 & E+\tfrac{1}{2}
		\end{array}
		\right).
	\end{align}
	The symplectic eigenvalues of $V^{(2)}$ are equal to $1$ and $1+4E$. The relative entropy for $K=2$ therefore evaluates to 
	\begin{align}
		D(\rho_E^{(2)}\|\rho_E^{\otimes 2}) 
		&= 2\cdot S(\rho_A) - S(\rho_{AB})\\
		&= 2\cdot g(E) - f(1+4\cdot E).
	\end{align}
	For $K>2$, the vectors 
	 $w_i:=-\mathbbm{i}e_1-e_2 -\mathbbm{i}e_i+e_{i+1}$for $i=3,5,\ldots,K-1$ are all eigenvectors to eigenvalue $1$ for $\mathbbm{i}\Omega V^{(K)}$, while $(\mathbbm{i},1,\mathbbm{i},1,\ldots,\mathbbm{i},1)$ is the vector to eigenvalue $(1+KE)$, as can be proven by induction.	
\begin{IEEEproof}[Proof of Theorem \ref{thm:quantum-advantage}]
	The series expansions follow from direct calculations, which we omit for brevity.
\end{IEEEproof}
\subsection{Linear optics and On-Off exponent}
We define the On-Off measurement as the POVM $\{\Pi_{0}, \Pi_{1}\}$ as $\Pi_{0} := \ket{0}\bra{0}$ and $\Pi_{1} := \mathds{1}-\ket{0}\bra{0}$, where one outcome corresponds to having no photons while the other one to a non-zero amount. The Hadamard transformation, $U_H$ \cite{Guha_2011}, is a linear transformation on $N=2^L$ modes of the electromagnetic field that transforms them according to 
\begin{align}
    U_{H}\textstyle\bigotimes_{j=1}^N\ket{\alpha_j} = \bigotimes_{i=1}^N \ket{\tfrac{1}{\sqrt{N}}\sum_j (-1)^{i \cross j}\alpha_j}
\end{align}
where $\cross$ denotes the bit-wise scalar product and $\ket{\alpha_j}$ are coherent states. Note that we only give its effect on coherent states, which is sufficient for our purposes. In particular, if $\alpha_j = \alpha \forall j$, then 
\begin{align}
    U_{H}\textstyle\bigotimes_{j=1}^N\ket{\alpha} = \ket{\sqrt{N}\alpha}\ket{0}^{\otimes N-1}.
\end{align}
That is, all the energy is focused on the first mode.  
We will calculate the measurement output distribution and then the relative entropy between distributions. 
    Given an incident field $|\alpha\rangle\langle\alpha|^{\otimes K}$ an optical Hadamard transformation will transform it to $|\sqrt{K}\alpha\rangle\langle\sqrt{K}\alpha|\otimes|0\rangle\langle0|^{\otimes (K-1)}$. Applying the POVM $\{\Pi_{b^K}\}_{b^K\in\{0,1\}^K}$ where $\Pi_{b^K}:=\otimes_{i=1}^K\Pi_{b_i}$ (On-Off measurement on each mode) and using  
    that for a coherent state $\bra{\alpha}\Pi_0 \ket{\alpha} = e^{-|\alpha|^2}$ it is clear that the probability distribution for On-Off measurement is a Bernoulli random variable with probabilities $e^{-|\alpha|^2}$ and $1-e^{-|\alpha|^2}$. Applying $U_H$ and calculating the probabilities of On-Off measurement on $\rho^{(K)}_E$ we find
\begin{align}
    U_{H}\rho^{(K)}_EU_H^{\dagger} = \tfrac{1}{\pi E}\int e^{-|\alpha|^2/E}U_H|\alpha\rangle\langle\alpha|^{\otimes K}U_H^{\dagger}d\alpha\\
    =\tfrac{1}{\pi E}\int e^{-|\alpha|^2/E} \ket{\sqrt{K}\alpha}\bra{\sqrt{K}\alpha}|0\rangle\langle 0|^{\otimes K -1}d\alpha
\end{align}
and when we apply the On-Off measurement on all modes, for the last $K-1$ modes the output will be 0 with probability 1, so we focus on the first mode. The probability of obtaining 0 is then given by 
\begin{align}
\label{eq : onoff}
    \tfrac{1}{\pi E}\int e^{-|\alpha|^2/E} \bra{\sqrt{K}\alpha}\Pi_1 \ket{\sqrt{K}\alpha}d\alpha =\\
    \tfrac{1}{\pi E}\int e^{-|\alpha|^2/E} e^{-K|\alpha|^2} d\alpha = \frac{1}{1+KE}.
\end{align}
The probability distribution is $P(\text{Off}) =\tfrac{1}{1+E}$ for the first mode and $P(\text{Off}) = 0$ for other modes, with $P(\text{On})$ being fixed by normalization. However, if the input state is $\rho_E^{\otimes K}$, then from calculations very similar to \autoref{eq : onoff}, we reach that each mode measurement output is a Bernouilli random variable $B(\tfrac{1}{1+E})$. We calculate the relative entropy by applying the classical discrete definition and again using that $D(r\otimes s|| u\otimes v) = D(r||u) + D(s||v)$ we obtain
\begin{equation}
    D_{\scriptscriptstyle{\mathrm{PH}}}(\rho^{(K)}_E, \rho^{\otimes K}_E) = d(E,K) + (K-1) \log(1+E).
\end{equation}

\subsection{Proofs of Lemma 4 and 5}
\begin{lemma}\label{lem:matrix-inverse}
	Let the $n\times n$ matrix $A$ have entries $(A)_{i,j}=(a-b)\delta_{i,j}+b$ where $a\neq b$. Then $A^{-1}$ exists and has entries 
	\begin{align}
		(A^{-1})_{i,j}=C^{-1}\cdot[(a+(n-2)b)\delta_{i,j} -b(1-\delta_{i,j})]
	\end{align}
	where $C=a^2 + (n-2)ab-(n-1)b^2$. Further, the determinant of $A$ is given by $\det(A)=(a-b)^{n-1}(a+(n-1)b)$.
\end{lemma}
\begin{IEEEproof}[Proof of Lemma \ref{lem:matrix-inverse}]
    The matrix $A$ has $n$ eigenvectors where the first is given by $\sum_{i=1}^{n}e_i$ with eigenvalue $a+(n-1)b$, and the next $n-1$ vectors are given by $-e_1+e_i$ (with eigenvalue $a-b$). As a result, the determinant is given by $(a-b)^{n-1}(a+(n-1)b)$.
\end{IEEEproof}

\begin{lemma}\label{lem:effective-dimension-for-phase-randomized-coherent-state}
	Let $\alpha\in\mathbb C$, $N\in\mathbb N$,  $T_N:=\sum_{n=0}^N|n\rangle\langle n|$. Then
	\begin{align}
		\tr T_N(|\alpha\rangle\langle\alpha|)\geq1 - \tfrac{1}{N!}\cdot  \max\{2,|\alpha|^2\}^{N}\label{eqn:performance-of-P_N-on-alpha}
	\end{align}
	Let $E>0$. If $N\geq E\cdot 50^3$, then
	\begin{align}
		\tr T_N\tfrac{1}{\pi E}\int e^{-|\alpha|^2/E}|\alpha\rangle\langle\alpha|d\alpha\geq 1-2^{- N}\label{eqn:performance-of-P_N-on-SE}.
	\end{align}
\end{lemma}
\begin{IEEEproof}[Proof of Lemma \ref{lem:effective-dimension-for-phase-randomized-coherent-state}]
	We first prove \eqref{eqn:performance-of-P_N-on-alpha}, from which \eqref{eqn:performance-of-P_N-on-SE} follows. To show this preliminary result, we first introduce the Gamma- and the incomplete Gamma functions:
	\begin{align}
		\Gamma(a)&:=\int_0^\infty t^{a-1}e^{-t}dt\\
		\gamma(a,z)&:=\int_0^z t^{a-1}e^{-t}dt\\
		\Gamma(a,z)&:=\int_z^\infty t^{a-1}e^{-t}dt.
	\end{align} 
	The incomplete gamma functions can be used to define a probability distribution \cite[6.5.1]{abramovitz}
	\begin{align}
		P(a,z):=\Gamma(a)^{-1}\gamma(a,z).
	\end{align}
	which \cite[6.5.13]{abramovitz} obeys the relation
	\begin{align}
		P(a,z)=1-e^{-z}\textstyle\sum_{n=0}^a\tfrac{z^n}{n!}
	\end{align}
	for $a\in\mathbb N$. Upon setting $z=|\alpha|^2$ and $a=N$ we obtain
	\begin{align}
		\tr T_N|\alpha\rangle\langle\alpha|=e^{-|\alpha|^2}\textstyle\sum_{n=0}^N\tfrac{|\alpha|^2}{N!}
		=1-\gamma(N,|\alpha|^2)/N!.\nonumber
	\end{align}
	To obtain a lower bound on $\tr T_N|\alpha\rangle\langle\alpha|$ we derive an upper bound on $\gamma(a,z)$ which is tight for $a\gg z>0$:
	\begin{align}
		\gamma(a,z) &= \int_0^1t^{a-1}e^{-t}dt + \int_1^zt^{a-1}e^{-t}dt\\
		&\leq \int_0^1t^{a-1}dt + \int_1^zt^{a-1}e^{-1}dt\\
		&= \tfrac{1}{a} + \tfrac{e^{-1}}{a}(z^a -1)\\
		&\leq\max\{2,z^a\}.
	\end{align}
	This shows the first part of Lemma \ref{lem:effective-dimension-for-phase-randomized-coherent-state}.
	We then utilize Stirling's approximation in the weak version $N!\geq N^Ne^{-N}$ and our assumption $N\geq8\cdot e\cdot\max\{2,|\alpha|^2\}$ to arrive at
	\begin{align}
		\tr T_N|\alpha\rangle\langle\alpha|&\geq1-4^{-N}. 
	\end{align}	
	Let us define $B(r):=\{\alpha:|\alpha|^2\leq r\}$. As a result of the formula $\tfrac{2}{E}\int_0^Rre^{-r^2/E}rdr = 1- e^{-R^2/E}$ we conclude that 
	\begin{align}
		\int &\tfrac{e^{-\tfrac{|\alpha|^2}{E}}}{\pi E}\tr T_N|\alpha\rangle\langle\alpha| d\alpha \geq \int_{B(N/50)} \tfrac{e^{-\tfrac{|\alpha|^2}{E}}}{\pi E}d\alpha(1-4^{-N})\nonumber\\
		&\geq (1-e^{-(N/50)^2/ E})(1-4^{-N})\geq1-2^{-N}.
	\end{align}
\end{IEEEproof}
\section{Conclusion}
    We studied a quantum hypothesis testing problem and used asymptotic exponents to quantify detection efficiencies. We found an unbounded quantum advantage in the low-energy regime. Our analysis shows that joint detection methods may not only improve communication systems \cite{Giovannetti2014,Banaszek_2020}, but also sensing abilities related to a variety of applications. Given that our problem statement describes the analysis of (weak) classical electromagnetic radiation, our findings shine some light on potential fields of application for quantum computing. More general models may be considered where phase or intensity differences between different detectors occur. Other concrete detection schemes \cite{Guha_2011,rosati2017decoding} and possible connections to recent work on learning classical structures from quantum data \cite{heidariLearningFromQuantumData} are also interesting open questions.

\bibliographystyle{plain}
\bibliography{bib}
\typeout{get arXiv to do 4 passes: Label(s) may have changed. Rerun}
\end{document}